\documentclass[aps,prd,reprint,onecolumn,10pt,eqsecnum,showpacs,nofootinbib,amsfonts,amssymb,amsmath,longbibliography,superscriptaddress,floatfix]{revtex4-1}
\usepackage[colorlinks=true,allcolors=blue, pdfusetitle]{hyperref}
\usepackage{bm}
\usepackage{mathrsfs}
\usepackage[dvipsnames]{xcolor}
\usepackage{mathtools}
\usepackage{graphicx}
\usepackage{orcidlink}
\usepackage{tensor}

\usepackage{algorithm}
\usepackage{algpseudocode}

\usepackage[commentmarkup=uwave]{changes}
\definechangesauthor[name=hs, color=cyan]{hs}

\usepackage{fontawesome5}

\usepackage{ragged2e}

\usepackage[T1]{fontenc}
\usepackage{lmodern}
\usepackage[utf8]{inputenc}
\usepackage{microtype}

\DeclareMathAlphabet{\mathbfsf}{\encodingdefault}{\sfdefault}{bx}{sl}

\usepackage[normalem]{ulem}

\begin{document}
	
\title{Thermodynamic properties and Joule-Thomson expansion of AdS black hole with Gaussian distribution in noncommutative geometry}

\author{Rui-Bo Wang\,\orcidlink{0009-0003-3198-3310}}
\email[Rui-Bo Wang: ]{wangrb2021@lzu.edu.cn}
\affiliation{School of Physical Science and Technology, Lanzhou University, Lanzhou, Gansu 730000, China}

\author{Lei You\,\orcidlink{0009-0004-5900-3318}}
\email[Lei You: ]{youl21@lzu.edu.cn}
\affiliation{School of Physical Science and Technology, Lanzhou University, Lanzhou, Gansu 730000, China}

\author{Shi-Jie Ma\,\orcidlink{0000-0003-4423-0142}}
\email[Shi-Jie Ma: ]{220220939731@lzu.edu.cn}
\affiliation{School of Physical Science and Technology, Lanzhou University, Lanzhou, Gansu 730000, China}

\author{Jian-Bo Deng\,\orcidlink{0000-0002-0586-6220}}
\email[Jian-Bo Deng(corresponding author): ]{dengjb@lzu.edu.cn}
\affiliation{School of Physical Science and Technology, Lanzhou University, Lanzhou, Gansu 730000, China}

\author{Xian-Ru Hu\,\orcidlink{0000-0001-9818-8635}}
\email[Xian-Ru Hu: ]{huxianru@lzu.edu.cn}
\affiliation{School of Physical Science and Technology, Lanzhou University, Lanzhou, Gansu 730000, China}

\date{\today}

\begin{abstract}
An investigation into the thermodynamics and Joule-Thomson expansion of anti-de Sitter black hole (AdS BH) with a Gaussian distribution in non-commutative geometry is presented. The metric of the Gaussian-distributed BH is derived, revealing a de Sitter (dS) geometry at its core. The results indicate that BH characterized by a Gaussian distribution exhibit thermodynamic properties remarkably similar to those of BH with a Lorentzian distribution in non-commutative geometry. This similarity is particularly manifested in the small-large BH phase transition, the modified first law of thermodynamics, critical behavior, heat capacity, the zeroth-order phase transition, and the Joule-Thomson process. Notably, the critical ratio of the Gaussian-distributed BH (0.46531) is significantly larger than that observed in Van der Waals fluid (0.375) and also substantially exceeds that of the Lorentzian-distributed BH (0.36671). Moreover, compared to the Lorentzian case, the zeroth-order phase transition effect in the Gaussian-distributed BH is exceedingly subtle—accompanied by a relative increase in the Gibbs free energy on the order of $10^{-3}\!\sim\!\!10^{-2}$—rendering it difficult to discern distinctly from the graph of the Gibbs free energy.

\hspace*{\fill}
	
Keywords: Noncommutative geometry, BH thermodynamics, Joule-Thomson expansion.

\end{abstract}

\maketitle
\section{Introduction}\label{Sect1}
Noncommutative geometry is a framework for quantized spacetime in which the coordinate operators no longer commute but satisfy the commutator $[\hat{x}^{\mu},\hat{x}^{\nu}]=\mathrm{i}\hat{\Theta}^{\mu\nu}$, where $\hat{x}^{\mu}$ denotes the spacetime coordinate operators and $\hat{\Theta}^{\mu\nu}$ is a nonzero anti-symmetric constant matrix. The elements of this matrix are denoted by $\Theta>0$, possessing the dimension of $\mathrm{L}^{2}$~\cite{th1,th2,th3,th4,th5,th6,th7}. The quantity $\sqrt{\Theta}$ is typically assumed to be of the order of the Planck length, $\ell_{p}\simeq1.616\times10^{-35}~\mathrm{m}$~\cite{commutator1,th1}, representing the minimal length scale of spacetime. Consequently, noncommutative effects become significant only at the extremely small Planck scale, whereas at macroscopic scales one may effectively take the limit \(\hbar\to0\), rendering the influence of noncommutative geometry negligible. Within the framework of general relativity, extensive research has been conducted to assess the impact of noncommutative geometry on gravitational theories~\cite{gr1,gr2,gr3,gr4,gr5,gr6,gr7,gr8}.

An intriguing consequence is that the nonzero anti-symmetric commutator between the spacetime coordinate operators violates Lorentz covariance~\cite{th1}, suggesting that noncommutative geometry might be intimately related to theories involving Lorentz symmetry breaking~\cite{lorentz1,lorentz2,lorentz3,lorentz4,lorentz5,lorentz6}. This observation offers a novel perspective for investigating Lorentz covariance violation. Research by A. Smailagic has demonstrated that noncommutative effects can eliminate the point-like mass distribution~\cite{th2,th3}, a phenomenon that may help avert the singularity problem in BHs. Inspired by this result, P. Nicolini and K. Nozari proposed replacing the point-mass distribution (represented by the Dirac delta function) with either a Gaussian distribution, $\rho_{G}=\frac{M}{\left(4\pi\Theta\right)^{\frac{3}{2}}}e^{-\frac{r^{2}}{4\Theta}}$~\cite{gauss1}, or a Lorentzian distribution, $\rho_{L}=\frac{M\sqrt{\Theta}}{\pi^{\frac{3}{2}}\left(r^{2}+\pi\Theta\right)^{2}}$~\cite{lorentz}, where $M$ denotes the mass of the point particle and $r$ is the radial coordinate in spherical coordinates. Both distributions reduce to the point-mass distribution $M\delta^{3}(x)$ (with $\delta^{3}(x)$ being the three-dimensional Dirac delta function) in the limit as the noncommutative parameter $\Theta$ tends to zero. Since these distribution functions remain finite and continuous at the core, the associated spacetime curvature does not diverge, thereby effectively preventing the formation of singularities. The properties of BHs within the framework of noncommutative geometry have attracted considerable attention~\cite{gauss2,gr1,gr2,gr3,gr4,gr9,gr10,gr11,gr12,gr13,effect,lorentz,optics,thermo1,thermo2,thermo3}. It has been shown, for instance, that even a neutral and nonrotating BH with a Gaussian mass distribution undergoes an evaporation process that culminates in an extremal BH with zero temperature, with the maximum temperature attained prior to reaching absolute zero remaining finite~\cite{gr1}. The influence of noncommutative geometry on the orbital dynamics of particles in central force fields has also been investigated~\cite{centralpotential}.

For Schwarzschild BH characterized by a Lorentzian distribution, extensive studies have been performed on aspects such as their shadows, gravitational lensing, quasinormal modes, and thermodynamics in extended phase space~\cite{optics,lorentz6,highlorentzian}. Compared to conventional Schwarzschild spacetimes, the Lorentzian distribution attenuates the gravitational effect, with its impact on the motion of timelike particles being significantly more pronounced than on the propagation of light~\cite{solartest}. Concerning the range of permissible values for the noncommutative parameter \(\Theta\), Silva constructed a gravitational vacuum star within the framework of Barnardos-Teitelboim-Zanelli theory under noncommutative geometry and obtained a parameter value of \(\sqrt{\Theta}\sim5.741\times10^{-21}~\mathrm{m}\)~\cite{BTZ}. Moreover, parameter constraints derived under the Lorentzian distribution have been meticulously determined from the four classical tests of general relativity within the solar system (namely, the precession of Mercury’s orbit, the deflection of light, radar echo delay, and gravitational redshift), with the precession of Mercury’s orbit yielding a relatively small parameter range \(\sqrt{\Theta}\leq0.25996~\mathrm{m}\)~\cite{solartest}. Nonetheless, both of these estimates for the order of magnitude of \(\sqrt{\Theta}\) are several orders of magnitude larger than the Planck length; obtaining more precise parameter values may ultimately require quantum mechanical experiments at microscopic scales. However, as another significant distribution in noncommutative geometry, the thermodynamic properties of BH characterized by a Gaussian distribution have not yet been systematically investigated.

BH thermodynamics is a relatively nascent branch of BH physics. Since the pioneering study by Hawking and Page on the phase transition in Schwarzschild-AdS BH in 1983~\cite{hawking}, the thermodynamics of BHs-especially within the extended phase space-has been extensively explored~\cite{xy1,xy2,xy3,xy4,xy5,xy6,xy7,xy8,xy9,xy10,xy11,xy12,xy13,xy14,xy15,xy16,xy17,xy18,xy19,xy20,xy21}. In 1999, Chamblin et al. discovered a first-order phase transition in Reissner-Nordström-AdS (RN-AdS) BH that is analogous to the gas-liquid phase transition~\cite{chamblin1,chamblin2}. In 2009, Kastor et al. proposed interpreting the cosmological constant as a pressure and derived the corresponding first law of BH thermodynamics~\cite{kastor}. In 2010, Dolan et al. computed the equation of state and identified the critical point for Kerr-Newman-AdS BH, revealing similarities with the Van der Waals fluid~\cite{xy17}. In 2012, David Kubizňák et al. were the first to identify $P-v$ critical phenomena in RN-AdS BH, systematically investigating the critical exponents and Gibbs free energy, and providing a detailed comparison with the thermodynamic properties of Van der Waals system~\cite{RNcri}. This work has established a mature methodology for studying BH thermodynamics in the extended phase space. In 2017, Özgür Ökcü et al. examined the Joule-Thomson process in RN-AdS BH, identifying a minimum inversion temperature and a minimum inversion mass, and compared their findings with those for Van der Waals fluid~\cite{JTRN}. As a typical representative of regular BHs, the thermodynamics of Bardeen-AdS BH have also attracted considerable attention~\cite{BD1,thermo2,BD3,BD4}. In summary, within the extended phase space, the cosmological constant in Einstein’s equations is regarded as an effective vacuum static ideal fluid, and it is assumed that the AdS BH reaches phase equilibrium with this fluid, thereby acquiring a pressure. This association of the cosmological constant with BH pressure not only addresses the incompleteness of the phase space in traditional BH thermodynamics but also leads to many intriguing research avenues. Moreover, within the context of the AdS/CFT correspondence, the interpretation of the cosmological constant as a thermodynamic variable has been revisited~\cite{BHChemistry,GE}.

In this paper, we investigate the thermodynamic properties and Joule-Thomson expansion of BH with a Gaussian distribution in noncommutative geometry. The Gaussian distribution represents an important alternative to the Lorentzian distribution in noncommutative geometry. However, the thermodynamic properties of BH characterized by this distribution have not been systematically and comprehensively investigated. We hope that this work will fill this gap and provide a meaningful comparison with existing studies on the thermodynamics of BH characterized by the Lorentzian distribution. Furthermore, this research may serve as a reference for future investigations.

This article is organized as follows. In Sect.~\ref{Sect2}, we derive the AdS BH with a Gaussian distribution within the framework of noncommutative geometry and discuss its minimal mass. In Sect.~\ref{Sect3}, we systematically investigate the thermodynamic properties of the BH in the extended phase space, including the modified first law, the small BH-large BH phase transition and criticality, critical exponents, isobaric heat capacity, Gibbs free energy, and the zeroth-order phase transition. In Sect.~\ref{Sect4}, we examine the Joule-Thomson expansion of the BH by plotting constant mass expansion curves and inversion curves, and analyzing the corresponding cooling and heating regions. Finally, in Sect.~\ref{Sect5} we summarize our study and provide an outlook for future research.

For convenience in mathematical formulation, natural units with \(\hbar=k_{B}=G=c=1\) are adopted throughout this paper.

\section{Schwarzschild-AdS BH with Gaussian distribution}\label{Sect2}
We commence our discussion from the Einstein equations incorporating the cosmological constant
\begin{equation}\label{eq2_1}
	R_{\mu}^{\nu}-\frac{1}{2}\delta_{\mu}^{\nu}R+\delta_{\mu}^{\nu}\Lambda=8\pi T_{\mu}^{\nu},
\end{equation}
where $R_{\mu}^{\nu}$ is the Ricci tensor, $\delta_{\mu}^{\nu}$ is the Kronecker symbol, $R$ is the scalar curvature and $T_{\mu}^{\nu}$ is the energy-momentum
tensor. One considers this following static spherically symmetric metric
\begin{equation}\label{eq2_2}
	ds^{2}=-f\left(r\right)dt^{2}+f\left(r\right)^{-1}dr^{2}+r^{2}d\theta^{2}+r^{2}\sin^{2}\theta d\phi^{2}.
\end{equation}
By substituting the Gaussian distribution~\cite{gauss1,gauss2,gr1}
\begin{equation}\label{eq2_3}
	\rho_{G}=\frac{M}{\left(4\pi\Theta\right)^{\frac{3}{2}}}e^{-\frac{r^{2}}{4\Theta}}
\end{equation}
into the Einstein equation, one could obtain
\begin{equation}\label{eq2_4}
	f\left(r\right)=1-\frac{2M}{r}\mathrm{Erf}\left(\frac{r}{2\sqrt{\Theta}}\right)+\frac{2Me^{-\frac{r^{2}}{4\Theta}}}{\sqrt{\pi\Theta}}-\frac{\Lambda r^{2}}{3},
\end{equation}
where $\mathrm{Erf}\left(x\right)$ is the error function
\begin{equation}\label{eq2_5}
	\mathrm{Erf}\left(x\right)=\frac{2}{\sqrt{\pi}}\int_{0}^{x}e^{-t^{2}}dt.
\end{equation}
For computational convenience, one could introduce a new variable
\begin{equation}\label{eq2_6}
	a=2\sqrt{\Theta}.
\end{equation}
Then Schwarzschild-AdS BH with Gaussian distribution in noncommutative geometry is finally derived
\begin{equation}\label{eq2_7}
	f\left(r\right)=1-\frac{2M}{r}\mathrm{Erf}\left(\frac{r}{a}\right)+\frac{4Me^{-\frac{r^{2}}{a^{2}}}}{a\sqrt{\pi}}-\frac{\Lambda r^{2}}{3}.
\end{equation}
When $r\to\infty$, this spacetime asymptotically approaches Schwarzschild-AdS spacetime.
\begin{equation}\label{eq2_8}
	f\left(r\right)\sim1-\frac{2M}{r}-\frac{\Lambda r^{2}}{3}.
\end{equation}
In particular, when $r\to0$,
\begin{equation}\label{eq2_9}
	f\left(r\right)\sim1-\frac{8Mr^{2}}{3a^{3}\sqrt{\pi}}-\frac{\Lambda r^{2}}{3}.
\end{equation}
It is evident that the Gaussian distribution smooths out the singularity, inducing a dS geometry at the core of the BH, characterized by an effective cosmological constant given by
\begin{equation}\label{eq2_10}
	\Lambda_{\mathrm{eff}}=\frac{8M}{a^{3}\sqrt{\pi}}.
\end{equation}
To ensure the existence of BH, it is necessary to require that $f\left(r_{h}\right)$ has solutions, with $r_{h}$ being the radius of event horizon. One could obtain this constraint
\begin{equation}\label{eq2_11}
	M>0.95206a.
\end{equation}

\section{Thermodynamics}\label{Sect3}

\subsection{The corrected first law}\label{Sect3_1}

The mass of the BH can be solved from equation $f\left(r_{h}\right)=0$
\begin{equation}\label{eq3_1}
	M=\frac{\sqrt{\pi}\xi e^{\xi^{2}}\left(3-\lambda \xi^{2}\right)}{6\left(-2\xi+\sqrt{\pi}e^{\xi^{2}}\mathrm{Erf}\left(\xi\right)\right)}a,
\end{equation}
where
\begin{equation}\label{eq3_2}
	r_{h}=\xi a,\qquad \Lambda=\frac{\lambda}{a^{2}}.
\end{equation}
It is worth noting that
\begin{equation}\label{eq3_3}
	\frac{\sqrt{\pi}\xi e^{\xi^{2}}\left(3-\lambda \xi^{2}\right)}{6\left(-2\xi+\sqrt{\pi}e^{\xi^{2}}\mathrm{Erf}\left(\xi\right)\right)}>0
\end{equation}
always holds true for $\lambda<0$ and $\xi>0$. Therefore, AdS BH characterized by a Gaussian distribution no longer possesses a minimum horizon radius.

The temperature of the BH is given by
\begin{equation}\label{eq3_4}
	T=\frac{f'\left(r_{h}\right)}{4\pi}=\frac{-6\xi+6\left(\lambda-2\right)\xi^{3}+4\lambda\xi^{5}-3\sqrt{\pi}e^{\xi^{2}}\left(\lambda\xi^{2}-1\right)\mathrm{Erf}\left(\xi\right)}{12a\pi\xi\left(\sqrt{\pi}e^{\xi^{2}}\mathrm{Erf}\left(\xi\right)-2\xi\right)}.
\end{equation}
In extended phase space, the cosmological constant is related to BH's pressure
\begin{equation}\label{eq3_5}
	P=-\frac{\Lambda}{8\pi}=-\frac{\lambda}{8\pi a^{2}}
\end{equation}
The entropy of the BH is given by the Bekenstein-Hawking formula
\begin{equation}\label{eq3_6}
	S=\frac{A}{4}=\pi r_{h}^{2},
\end{equation}
where $A=4\pi r_{h}^{2}$ is the horizon area.

It is shown in Refs.~\cite{BD3,modify1,modify2,modify3,lorentzianBH} that when the energy-momentum tensor outside the horizon is related to the mass of the BH, the first law takes a corrected form
\begin{equation}\label{eq3_7}
	WdM=TdS+VdP,
\end{equation}
where $W$ is correction function, whose general form takes
\begin{equation}\label{eq3_8}
	W=1-\frac{\partial}{\partial M}\int_{r_{h}}^{\infty}\rho d^{3}x,
\end{equation}
where the integration region extends throughout the exterior of the BH horizon. For the BH we are discussing,
\begin{equation}\label{eq3_9}
	W=1-\frac{2\xi e^{-\xi^{2}}}{\sqrt{\pi}}-\mathrm{Erfc}\left(\xi\right),
\end{equation}
where
\begin{equation}\label{eq3_10}
	\mathrm{Erfc}\left(\xi\right)=1-\mathrm{Erf}\left(\xi\right)=\frac{2}{\sqrt{\pi}}\int_{\xi}^{\infty}e^{-t^{2}}dt.
\end{equation}
Specifically, when $a\to0$, it is equivalent that $\xi\to\infty$, which leads to $W\to1$ that describes the normal first law. 

One could verify
\begin{equation}\label{eq3_11}
	T=W\left(\frac{\partial M}{\partial S}\right)_{P}.
\end{equation}
The thermodynamic volume of the BH is
\begin{equation}\label{eq3_12}
	V=W\left(\frac{\partial M}{\partial P}\right)_{S}=\frac{4\pi r_{h}^{3}}{3}.
\end{equation}

\subsection{Phase transition and criticality}\label{Sect3_2}
\begin{figure}[htbp]
\centering
\begin{minipage}{0.49\textwidth}
	\centering
	\includegraphics[width=\linewidth]{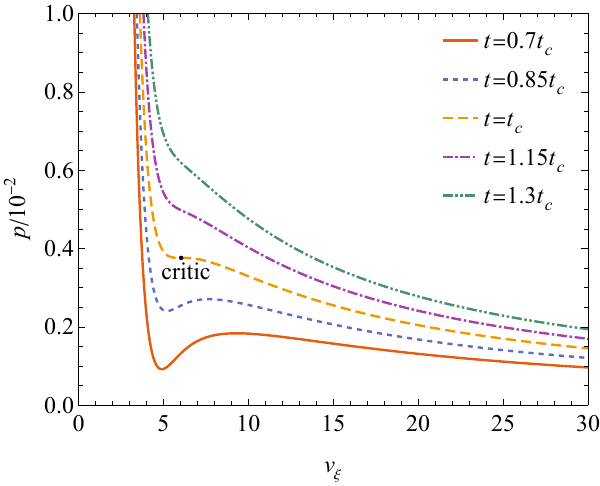}
	\caption{The figure of $p\left(v_\xi\right)$ of AdS BH with Gaussian distribution. We set $t=0.7t_{c}, 0.85t_{c}, t_{c}, 1.15t_{c}, 1.3t_{c}$ respectively. The point "critic" is the critical point.}\label{Pv}
\end{minipage}
\hfill
\begin{minipage}{0.49\textwidth}
	\centering
	\includegraphics[width=\linewidth]{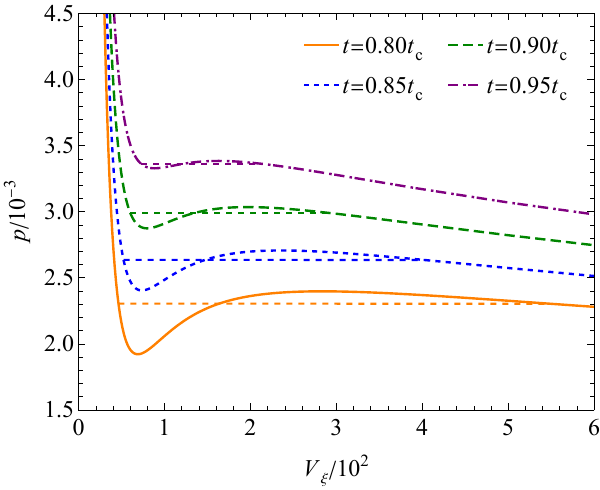}
	\caption{Maxwell's equal area law in the $p-V_{\xi}$ coordinate system, the horizontal dashed lines indicate the phase transition. We set $t=0.8t_{c}, 0.85t_{c}, 0.9t_{c}, 0.95t_{c}$ respectively.}\label{pvmaxwell}
\end{minipage}
\end{figure}

\begin{figure}[htbp]
\centering
\begin{minipage}{0.49\textwidth}
	\centering
	\includegraphics[width=\linewidth]{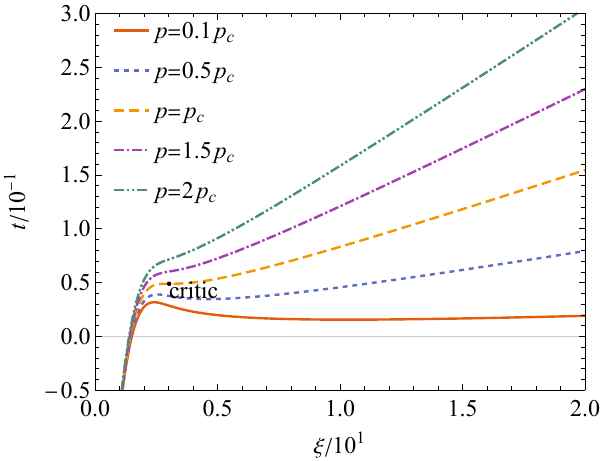}
	\caption{The figure of $t\left(\xi\right)$ of AdS BH with Gaussian distribution. We set $p=0.1p_{c}, 0.5p_{c}, p_{c}, 1.5p_{c}, 2p_{c}$ respectively. The point "critic" is the critical point.}\label{txi}
\end{minipage}
\hfill
\begin{minipage}{0.49\textwidth}
	\centering
	\includegraphics[width=\linewidth]{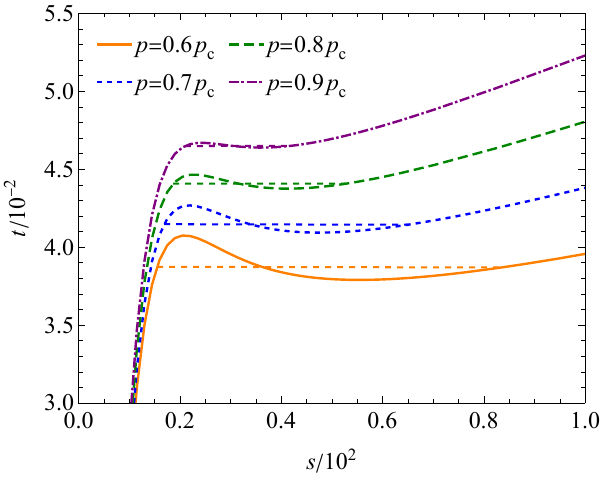}
	\caption{Maxwell's equal area law in the $t-s$ coordinate system, the horizontal dashed lines indicate the phase transition. We set $p=0.6p_{c}, 0.7p_{c}, 0.8p_{c}, 0.9p_{c}$ respectively.}\label{tsmaxwell}
\end{minipage}
\end{figure}

The equation of state of the BH could be obtained from Eq.~\ref{eq3_4}
\begin{equation}\label{eq3_13}
	p=-\frac{1}{8\pi\xi^{2}}+\frac{t}{2\xi}-\frac{\xi\left(2\pi t\xi+1\right)}{\pi(6\xi+4\xi^{3}-3\sqrt{\pi}e^{\xi^{2}}\mathrm{Erf}\left(\xi\right))},
\end{equation}
where
\begin{equation}\label{eq3_14}
	p=-\frac{\lambda}{8\pi}=a^{2}P,\quad t=aT.
\end{equation}
The critical point satisfies
\begin{equation}\label{eq3_15}
	\frac{\partial p}{\partial\xi}=\frac{\partial^{2}p}{\partial\xi^{2}}=0,
\end{equation}
which has solution
\begin{equation}\label{eq3_16}
	\xi_{c}=3.0219,\quad t_{c}=0.048823,\quad p_{c}=0.0037589.
\end{equation}
So the critical point is derived
\begin{equation}\label{eq3_17}
	P_{c}=\frac{0.0037589}{a^{2}},\quad r_{c}=3.0219a,\quad T_{c}=\frac{0.048823}{a^{2}}.
\end{equation}
Introducing the specific volume $v=2r_{h}$~\cite{v1,v3,v4,RNcri}, one has critical ratio
\begin{equation}\label{eq3_18}
	\frac{P_{c}v_{c}}{T_{c}}=0.46531.
\end{equation}
Compared with the result of Lorentzian distribution $0.36671$~\cite{lorentzianBH}, this numerical value significantly deviates from $0.375$. One could also introduce $v_{\xi}=2\xi=v/a$ and plot the BH isotherms. Fig.~\ref{Pv} illustrates the $p-v_{\xi}$ critical phenomena of BH. It is evident that the $P-v$ curve for Gaussian-distributed AdS BH is remarkably analogous to that of a Van der Waals system. For temperatures below the critical temperature ($t<t_{c}$), an unstable region emerges in which the pressure increases with volume rather than decreases, indicating a phase transition between small BH-large BH. By analogy with the Van der Waals fluid, this small BH-large BH phase transition can be described using Maxwell's equal area law~\cite{RNcri,modify2,ratio1,ratio2}
\begin{equation}\label{eq3_19}
	\oint VdP=0.
\end{equation}
The Maxwell's equal area law in $P-V$ coordinate system is plotted in Fig.~\ref{pvmaxwell}. In the figure, we make a non-dimensionalization
\begin{equation}\label{eq3_20}
	V_{\xi}=\frac{V}{a^{3}}=\frac{4\pi\xi^{3}}{3}.
\end{equation}
The critical behavior of BHs can also be manifested in the temperature-entropy diagram, as shown in Fig.~\ref{txi}. In this figure, in order to preserve generality, we also introduce a non-dimensionalized entropy
\begin{equation}\label{eq3_21}
	s=\frac{S}{a^{2}}=\pi\xi^{2}.
\end{equation}
Similarly, Maxwell's equal area law holds in the $t-s$ coordinate system~\cite{ratio3}
\begin{equation}\label{eq3_22}
	\oint SdT=0,
\end{equation}
as depicted in Fig.~\ref{tsmaxwell}. This result will be instrumental in our subsequent analysis of the Gibbs free energy of BH.

\subsection{Critical exponents}\label{Sect3_3}
By introducing dimensionless variables
\begin{equation}\label{eq3_23}
	p'=\frac{p}{p_{c}},\quad t'=\frac{t}{t_{c}},\quad\xi'=\frac{\xi}{\xi_{c}},
\end{equation}
one could get the equation of corresponding state
\begin{equation}\label{eq3_24}
	p'=-\frac{1}{8\pi p_{c}\xi_{c}^{2}\xi'^{2}}+\frac{t_{c}t'}{2p_{c}\xi_{c}\xi'}+\frac{\xi_{c}\xi'\left(2\pi t_{c}\xi_{c}t'\xi'+1\right)}{\pi(6\xi_{c}\xi'+4\xi_{c}^{3}\xi'^{3}-3\sqrt{\pi}e^{\xi_{c}^{2}\xi'^{2}}\mathrm{Erf}\left(\xi_{c}\xi'\right))}.
\end{equation}
To calculate the critical exponents, one could perform a transformation
\begin{equation}\label{eq3_25}
	\tau=t'-1,\quad w=\frac{V}{V_{c}}-1=\xi'^{3}-1,
\end{equation}
and expand the function $p'$ as series
\begin{equation}\label{eq3_26}
	p=1+c_{10}\tau-c_{11}\tau w+c_{12}\tau w^{2}-c_{03}w^{3}+\mathcal{O}\left(\tau w^{3},w^{4}\right).
\end{equation}
where $c_{10}$, $c_{11}$, $c_{12}$, and $c_{03}$ are both positive constants.

The critical exponents are defined by~\cite{exponentdef}
\begin{align}
	&C_{V}\propto\left|\tau\right|^{-\alpha},\label{eq3_27}\\
	&V_{l}-V_{s}\propto\left|\tau\right|^{\beta},\label{eq3_28}\\
	&\kappa_{T}=-\frac{1}{V}\left(\frac{\partial V}{\partial P}\right)_{T}\propto\left|\tau\right|^{-\gamma},\label{eq3_29}\\
	&\left|P-P_{c}\right|\propto\left|V-V_{c}\right|^{\delta}\label{eq3_30},
\end{align}
where $C_{V}$ is the isochoric heat capacity. $V_{l}-V_{s}$ is the phase transition volume. $\kappa_{T}$ is the isothermal compressibility. The fourth equation describes the behavior of $P-V$ graph at critical temperature.

From Eqs.~\ref{eq3_6} and \ref{eq3_12}, it is obvious that
\begin{equation}\label{eq3_31}
	dV=0\Rightarrow dS=0\Rightarrow C_{V}=T\left(\frac{\partial S}{\partial T}\right)_{V}=0,
\end{equation}
which gives $\alpha=0$.

In order to calculate the phase transition volume, one should use Maxwell's area law
\begin{equation}\label{eq3_32}
	p'\left(w_{s}\right)=p'\left(w_{l}\right),\qquad\oint\left(w+1\right)dp'=0.
\end{equation}
One could obtain
\begin{equation}\label{eq3_33}
\begin{aligned}
	\Delta w&=w_{l}-w_{s}=\frac{2\sqrt{\tau\left(c_{12}^{2}\tau-3c_{11}c_{03}\right)}}{\sqrt{3}c_{03}}\\
	&=2\sqrt{\frac{-c_{11}\tau}{c_{03}}}+\mathcal{O}\left(-\tau\right)^{\frac{3}{2}},
\end{aligned}
\end{equation}
which gives $\beta=\frac{1}{2}$.

On the other hand,
\begin{equation}\label{eq3_34}
	\kappa_{T}\propto\frac{1}{\left(w+1\right)}\frac{dw}{dp'}=-\frac{1}{c_{11}\tau}+\mathcal{O}\left(w\right),
\end{equation}
which yields $\gamma=1$.

Finally, when $t=0$,
\begin{equation}\label{eq3_35}
	p'=1-c_{03}w^{3},
\end{equation}
which causes $\delta=3$.

It can be observed that the critical exponents of BH with a Gaussian distribution are identical to those with a Lorentzian distribution~\cite{lorentzianBH}. Moreover, the critical exponents of these two types of BH coincide with those of the Van der Waals system, and they both satisfy the Griffiths, Rushbrooke and Widom formulas~\cite{ratio1,exponentdef,exponent}
\begin{align}
	&\alpha+\beta\left(\delta+1\right)=2,&\text{Griffiths}\label{eq3_36}\\
	&\gamma\left(\delta+1\right)=\left(2-\alpha\right)\left(\delta-1\right),&\text{Griffiths}\label{eq3_37}\\
	&\alpha+2\beta+\gamma=2,&\text{Rushbrooke}\label{eq3_38}\\
	&\gamma=\beta\left(\delta-1\right).&\text{Widom}\label{eq3_39}
\end{align}

\subsection{Heat capacity}\label{Sect3_4}
\begin{figure}[htbp]
	\centering
	\includegraphics[width=1\textwidth]{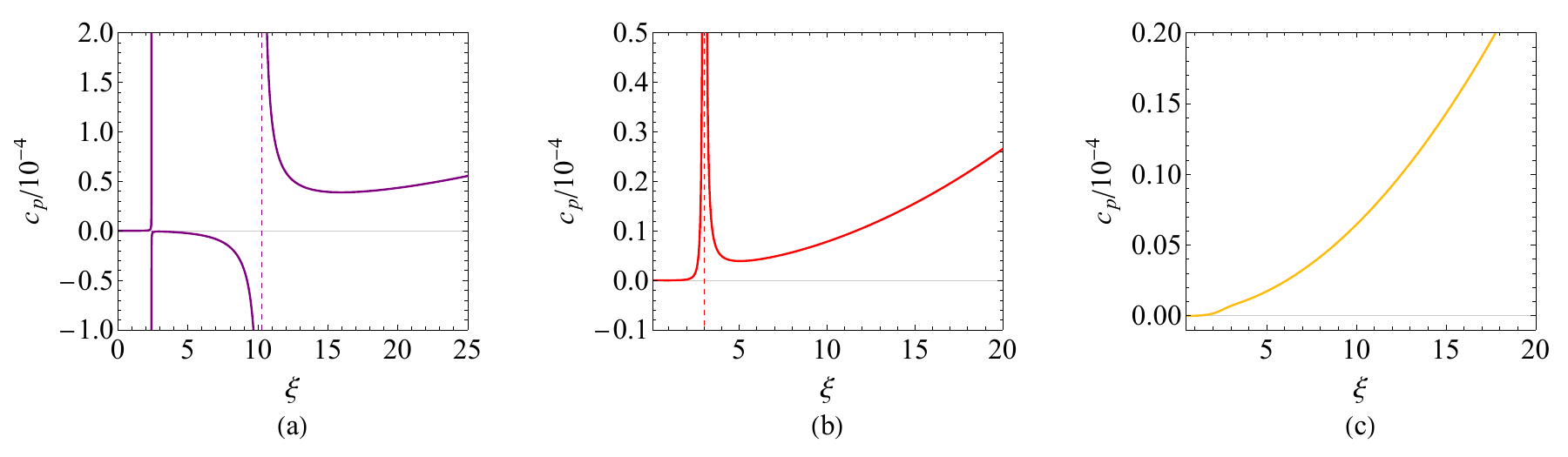}
	\caption{The heat capacity $c_{p}$ as a function of $\xi$. We set (a)$p=0.1p_{c}$; (b)$p=p_{c}$; (c)$p=10p_{c}$.}\label{cp}
\end{figure}

The isobaric heat capacity of the BH is
\begin{equation}\label{eq3_40}
\begin{aligned}
	c_{p}&=t\left(\frac{\partial s}{\partial t}\right)_{p}\\
	&=\frac{2\pi\xi^2\left(e^{\xi^2}\sqrt{\pi}\operatorname{Erf}(\xi)-2\xi\right)\left(6\xi-6(\lambda-2)\xi^3-4\lambda\xi^5+3e^{\xi^2}\sqrt{\pi}(\lambda\xi^2-1)\operatorname{Erf}(\xi)\right)}{4\xi^2\left(3+3(\lambda-4)\xi^2+8\lambda\xi^4\right)+4e^{\xi^2}\sqrt{\pi}\xi\left(-3-3(\lambda-2)\xi^2-2(3+2\lambda)\xi^4+2\lambda\xi^6\right)\operatorname{Erf}(\xi)+3e^{2\xi^2}\pi(1+\lambda\xi^2)\operatorname{Erf}(\xi)^2}.
\end{aligned}
\end{equation}
Fig.~\ref{cp} illustrates the isobaric heat capacity of BH for temperatures below, equal to, and above the critical temperature $t_{c}$. It is evident that the behavior of the heat capacity corresponds well with the BH's temperature-entropy diagram. At the point where $dt/ds=0$, the heat capacity diverges to infinity, indicating a phase transition in the BH system. In the unstable region where $dt/ds<0$ (see Fig.~\ref{txi}), the heat capacity assumes negative values. In particular, when the event horizon of the BH is extremely small $\xi<\xi_{min}$, where $\xi_{min}$ is the positive root of equation
\begin{equation}\label{eq3_41}
	6\xi-6(\lambda-2)\xi^{3}-4\lambda\xi^{5}+3\sqrt{\pi}e^{\xi^{2}}(\lambda\xi^{2}-1)\mathrm{Erf}(\xi)=0.
\end{equation}
the temperature becomes negative (see Fig.~\ref{txi}), which in turn leads to a negative heat capacity. So a Gaussian-distributed AdS BH with a sufficiently small horizon radius is unstable. This unique property is also exhibited by other BHs~\cite{BD1,modify3,lorentzianBH}.

Furthermore, the heat capacity exhibits asymptotic behavior for both very small and very large horizon radii
\begin{equation}\label{eq3_42}
	\lim_{\xi\to0}c_{p}=0,\quad\lim_{\xi\to\infty}c_{p}=\infty.
\end{equation}
Notably, the isobaric heat capacity behavior of BH with a Gaussian distribution is entirely consistent with that of BH with a Lorentzian distribution.

\begin{figure}[htbp]
	\centering
	\includegraphics[width=0.49\textwidth]{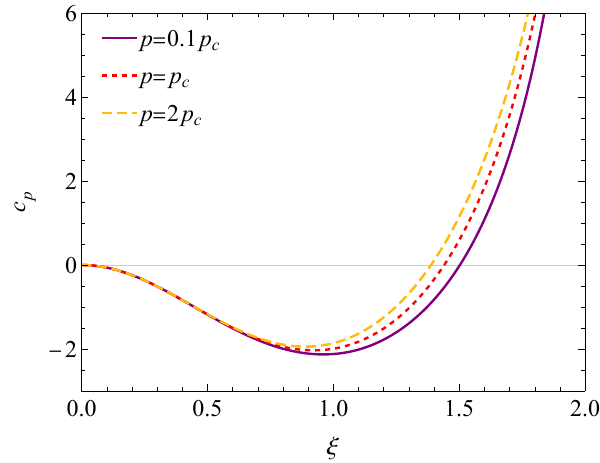}
	\caption{The negative section of the heat capacity $c_{p}$. We set $p=0.1p_{c}, p_{c}, 2p_{c}$.}\label{cpn}
\end{figure}

\subsection{Gibbs free energy}\label{Sect3_5}
The Gibbs free energy is defined as
\begin{equation}\label{eq3_43}
	G=M-TS=ga.
\end{equation}
So one has
\begin{equation}\label{eq3_44}
	g=m-ts=\frac{\xi\left(-6\xi-4\xi^3\left(3+4p\pi(3+2\xi^2)\right)+e^{\xi^2}\sqrt{\pi}\left(8p\pi\xi^2-3-3\left(1+8p\pi\xi^2\right)\operatorname{Erfc}(\xi)\right)\right)}{24\xi-12e^{\xi^2}\sqrt{\pi}\operatorname{Erf}(\xi)},
\end{equation}
where $M=ma$.

\begin{figure}[htbp]
\centering
\begin{minipage}{0.51\textwidth}
	\centering
	\includegraphics[width=\linewidth]{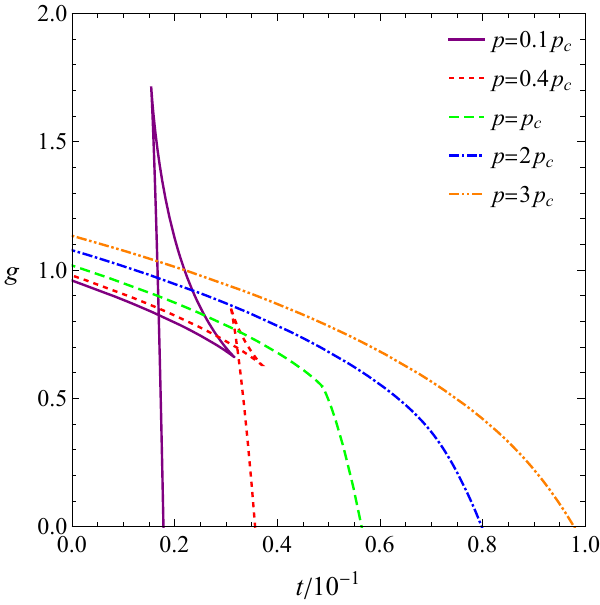}
	\caption{The Gibbs free energy of the AdS BH with Gaussian distribution as a function of BH temperature. We set $p=0.1p_{c}, 0.4p_{c}, p_{c}, 2p_{c}, 3p_{c}$ respectively.}\label{glarge}
\end{minipage}
\hfill
\begin{minipage}{0.48\textwidth}
	\centering
	\includegraphics[width=\linewidth]{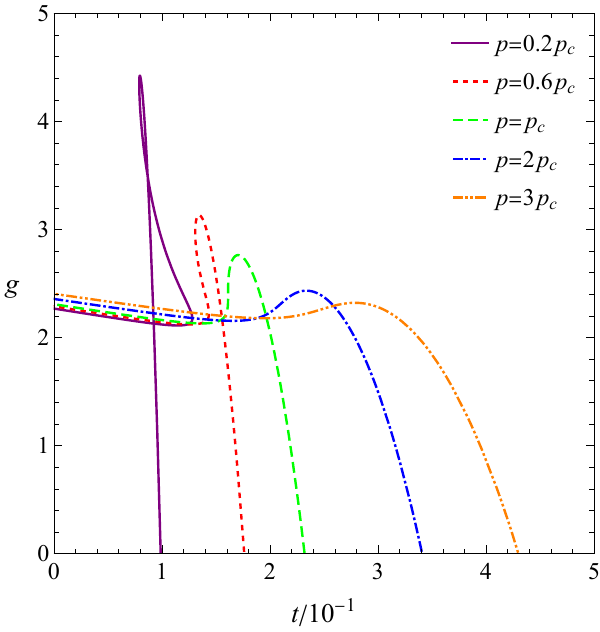}
	\caption{The Gibbs free energy of the Lorentzian-distributed BH. For the sake of comparison, we also set $2\sqrt{\Theta}=1$. We set $p=0.2p_{c}, 0.6p_{c}, p_{c}, 2p_{c}, 3p_{c}$, where $p_{c}$	represents the critical pressure of the Lorentzian-distributed BH.}\label{log}
\end{minipage}
\end{figure}

Fig.~\ref{glarge} displays the Gibbs free energy of the BH.

It is apparent from the figure that the Gibbs free energy behavior closely resembles that of a Van der Waals fluid. However, according to the studies reported in~\cite{modify3,lorentzianBH}, for BHs described by the corrected first law, the small BH-large BH phase transition is a zeroth-order phase transition, characterized by a discontinuity in the Gibbs free energy. This phenomenon is particularly pronounced in BH with a Lorentzian distribution in noncommutative geometry~\cite{lorentzianBH}. The Gibbs free energy of Lorentzian-distributed BH is shown in Fig.~\ref{log}. The figure clearly demonstrates that even when the BH pressure is below the critical pressure, the Gibbs free energy does not necessarily exhibit self-intersection. Moreover, the Gibbs free energy is not always monotonically decreasing with temperature. These behaviors significantly differ from those of a Van der Waals fluid.

In contrast, we will demonstrate that for BH with a Gaussian distribution, the zeroth-order phase transition effect is considerably weakened, making it less readily apparent in the $g\left(t\right)$ diagram.

In order to further calculate the zeroth-order phase transition, one could start from Eq.~\ref{eq3_43}
\begin{equation}\label{eq3_45}
	dg=\left(W^{-1}-1\right)tds-sdt+W^{-1}V_{\xi}dp.
\end{equation}
Considering the Maxwell's equal area law in $t-s$ coordinate system
\begin{equation}\label{eq3_46}
	\oint sdt=0,
\end{equation}
the Gibbs free energy increment for the small BH-large BH process is given by
\begin{equation}\label{eq3_47}
	\Delta g=\int_{s_{s}}^{s_{l}}\left(W^{-1}-1\right)tds.
\end{equation}
Taking the condition $p=0.9p_{c}$ for an example, one could obtain the phase transition location
\begin{equation}\label{eq3_48}
	\xi_{s}=2.5962,\quad\xi_{l}=3.6664,
\end{equation}
which yields a relative increase rate in the Gibbs free energy
\begin{equation}\label{eq3_49}
	\frac{\Delta g}{g(\xi_{s})}=\frac{g(\xi_{l})-g(\xi_{s})}{g(\xi_{s})}=1.0\times10^{-3}\ll1.
\end{equation}
So the zeroth-order phase transition effect is not pronounced and can be considered nearly negligible.

Here, we present the relative growth amplitude of the zeroth-order phase transition in noncommutative geometry AdS BHs under certain pressures (see Tab.~\ref{zeropt}). It is evident that the zeroth-order phase transition effect is significantly pronounced for BH with a Lorentzian distribution, resulting in a Gibbs free energy curve that distinctly deviates from that of a Van der Waals fluid. In contrast, for BH with a Gaussian distribution, the relative growth amplitude of the Gibbs free energy during the small BH-large BH phase transition is considerably smaller, on the order of $10^{-3}$ to $10^{-2}$. Consequently, the zeroth-order phase transition effect in Gaussian-distributed BH is extremely weak, making it difficult to directly discern from the Gibbs free energy curve.

\begin{table}[htbp]
\centering
\caption{The Relative increase in Gibbs free energy during the zeroth-order phase transition of AdS BHs in noncommutative geometry at different pressures. The second row corresponds to the Gaussian distribution, while the third row represents the Lorentzian distribution.}
\setlength{\tabcolsep}{1.26mm}{
	\begin{tabular}{cccccccccccc}
	\hline\hline
	$p/p_{c}$&0.9&0.8&0.7&0.6&0.5&0.4&0.3&0.2&$10^{-1}$&$10^{-2}$&$10^{-3}$\\
	\hline
	$\Delta g_{G}/g_{G}$&$1.0\!\!\times\!\!10^{-3}$&$2.0\!\!\times\!\!10^{-3}$&$3.0\!\!\times\!\!10^{-3}$&$4.3\!\!\times\!\!10^{-3}$&$5.7\!\!\times\!\!10^{-3}$&$7.2\!\!\times\!\!10^{-3}$&$8.9\!\!\times\!\!10^{-3}$&$1.1\!\!\times\!\!10^{-2}$&$1.3\!\!\times\!\!10^{-2}$&$1.4\!\!\times\!\!10^{-2}$&$1.4\!\!\times\!\!10^{-2}$\\
	\hline
	$\Delta g_{L}/g_{L}$&0.21&0.31&0.39&0.47&0.55&0.63&0.72&0.82&0.96~~~&1.3~~~&1.6~~~\\
	\hline\hline
	\end{tabular}}
	\label{zeropt}
\end{table}

\section{Joule-Thomson expansion}\label{Sect4}
The Joule-Thomson expansion of BHs is an intriguing topic that has garnered extensive discussion within the academic community~\cite{JTRN,JT1,JT2,JT3,JT4,JT5,JT6,JT7,JT8,JT9,JT10,JT11,JT12,JT13,JT14,JT15,JT16,JT17,JT18}. The Joule-Thomson process is an isenthalpic process. Considering the mass of AdS BH is regarded as its enthalpy, the BH's Joule-Thomson process precisely corresponds to an expansion process at constant mass
\begin{equation}\label{eq4_1}
	dM=0.
\end{equation}
The Joule-Thomson coefficient is
\begin{equation}\label{eq4_2}
	\mu_{\mathrm{JT}}=\left(\frac{\partial T}{\partial P}\right)_{M}.
\end{equation}
For computational convenience, both temperature $t$ and pressure $p$ can be expressed explicitly as functions of mass $m$ and horizon radius $\xi$:
\begin{equation}\label{eq4_3}
	p=\frac{3\left(-4e^{-\xi^2}m\xi-\sqrt{\pi}\xi+2m\sqrt{\pi}\mathrm{Erf}(\xi)\right)}{8\pi^{3/2}\xi^{3}},
\end{equation}
\begin{equation}\label{eq4_4}
	t=\frac{-2e^{-\xi^2}m\xi(3+2\xi^2)-\sqrt{\pi}\left(\xi-3m\mathrm{Erf}(\xi)\right)}
	{2\pi^{3/2}\xi^2}.
\end{equation}
So one could have
\begin{equation}\label{eq4_5}
	\mu_{\mathrm{JT}}=\left(\frac{\partial t}{\partial p}\right)_{m}=\frac{\left(\partial t/\partial\xi\right)_{m}}{\left(\partial p/\partial\xi\right)_{m}}
	=\frac{2\xi\left(4m\xi\left(3+2\left(\xi^2+\xi^4\right)\right)+e^{\xi^2}\sqrt{\pi}\left(\xi-6m\mathrm{Erf}(\xi)\right)\right)}{6m\xi\left(3+2\xi^2\right)+3e^{\xi^2}\sqrt{\pi}\left(\xi-3m\mathrm{Erf}(\xi)\right)}.
\end{equation}

By setting $\mu_{\mathrm{JT}}=0$, one gets inversion mass
\begin{equation}\label{eq4_6}
	m_{i}=\frac{-e^{\xi^2}\sqrt{\pi}\xi}{2\left(6\xi+4\xi^3+4\xi^5-3e^{\xi^2}\sqrt{\pi}\mathrm{Erf}(\xi)\right)}.
\end{equation}
By substituting this solution into the expression of $t$ and $p$, one acquires the inversion temperature $t_{i}$ and the inversion pressure $p_{i}$
\begin{equation}\label{eq4_7}
	p_{i}=\frac{6\left(\xi+\xi^3+\xi^5\right)-3e^{\xi^2}\sqrt{\pi}\operatorname{Erf}(\xi)}{4\xi^2\left(-2\pi\xi\left(3+2\xi^2+2\xi^4\right)+3e^{\xi^2}\pi^{3/2}\operatorname{Erf}(\xi)\right)},
\end{equation}
\begin{equation}\label{eq4_8}
	t_{i}=\frac{-6\xi-4\xi^3-8\xi^5+3e^{\xi^2}\sqrt{\pi}\operatorname{Erf}(\xi)}{24\pi\xi^2+16\pi\xi^4+16\pi\xi^6-12e^{\xi^2}\pi^{3/2}\xi\operatorname{Erf}(\xi)}.
\end{equation}
\begin{figure}[htbp]
\centering
\begin{minipage}{0.49\textwidth}
	\centering
	\includegraphics[width=\linewidth]{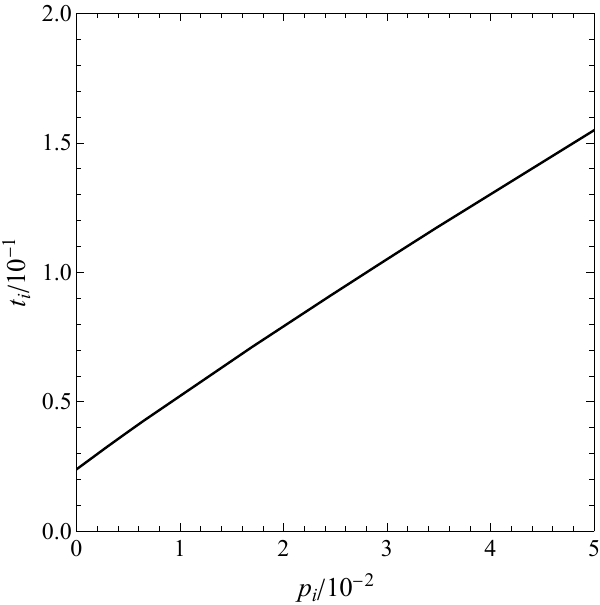}
	\caption{The inversion curve of AdS BH with Gaussian distribution.}\label{invnd}
\end{minipage}
\hfill
\begin{minipage}{0.49\textwidth}
	\centering
	\includegraphics[width=\linewidth]{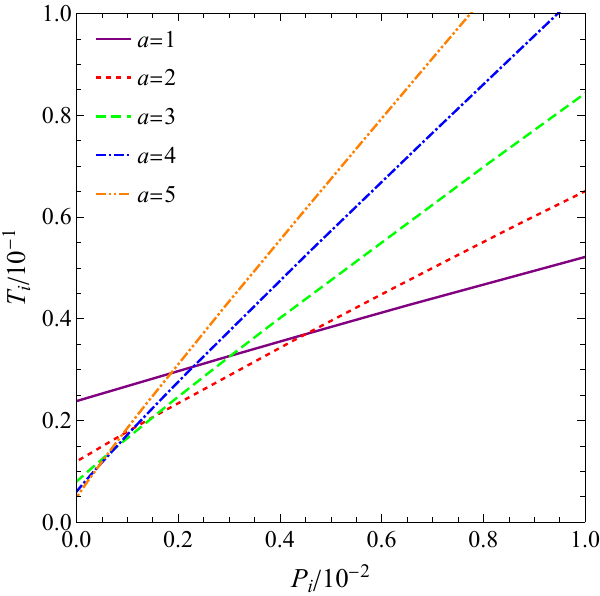}
	\caption{The inversion curves for different noncommutative parameters. We set $a=1, 2, 3, 4$, and $5$.}\label{inv}
\end{minipage}
\end{figure}

The BH's inversion curve is shown in Fig.~\ref{invnd}. Similar to the BH with Lorentzian distribution, the BH with Gaussian distribution doesn't exhibit a maximum inversion pressure or inversion temperature. In particular, there exists a minimum inversion pressure, $p_{i}=0$, at which the BH attains its minimum inversion temperature $t_{i}$. When $p_{i}=0$, $\xi=1.9193$. One have
\begin{equation}\label{eq4_9}
	t_{i}=0.023757,
\end{equation}
or
\begin{equation}\label{eq4_10}
	T_{i}=\frac{0.023757}{a}.
\end{equation}
It is worth noticing that this result contributes a dimensionless constant
\begin{equation}\label{eq4_11}
	\frac{T_{i}}{T_{c}}=0.48660.
\end{equation}

The inversion curves for different noncommutative parameters are shown in Fig.~\ref{inv}.

\begin{figure}[htbp]
	\centering
	\includegraphics[width=0.49\textwidth]{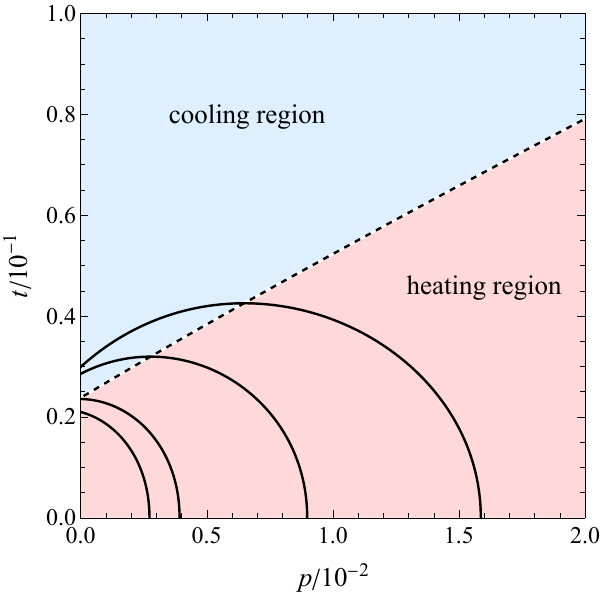}
	\caption{The constant mass expansion curves (solid lines) and inversion curves (dashed line) of BH. We set $a=1, m_{i}^{min}=1.022, 1.1, 1.2$. The red region represents heating region ($\mu_{\mathrm{JT}}<0$) and the blue part is cooling zone ($\mu_{\mathrm{JT}}>0$).}\label{hcf}
\end{figure}

We have also plotted the constant mass expansion curves and inversion curve of the BH (see Fig.~\ref{hcf}). It is evident that the inversion curve partitions the $T-P$ plane into two regions: the upper-left region corresponds to the cooling region (where $\mu_{\mathrm{JT}}>0$),  whereas the lower-right region corresponds to the heating region (where $\mu_{\mathrm{JT}}<0$).

Specifically, there exists a minimum inversion mass. By substituting $p_{i}=0$ ($\xi=1.9193$) into Eq.~\ref{eq4_6}, one gets minimum inversion mass
\begin{equation}\label{eq4_12}
	m_{i}^{min}=1.0221,
\end{equation}
or
\begin{equation}\label{eq4_13}
	M_{i}^{min}=1.0221a.
\end{equation}
Any BH with a mass less than $M_{i}^{min}$ cannot experience an inversion point; consequently, it remains indefinitely in the heating regime.

\begin{figure}[htbp]
	\centering
	\includegraphics[width=0.85\textwidth]{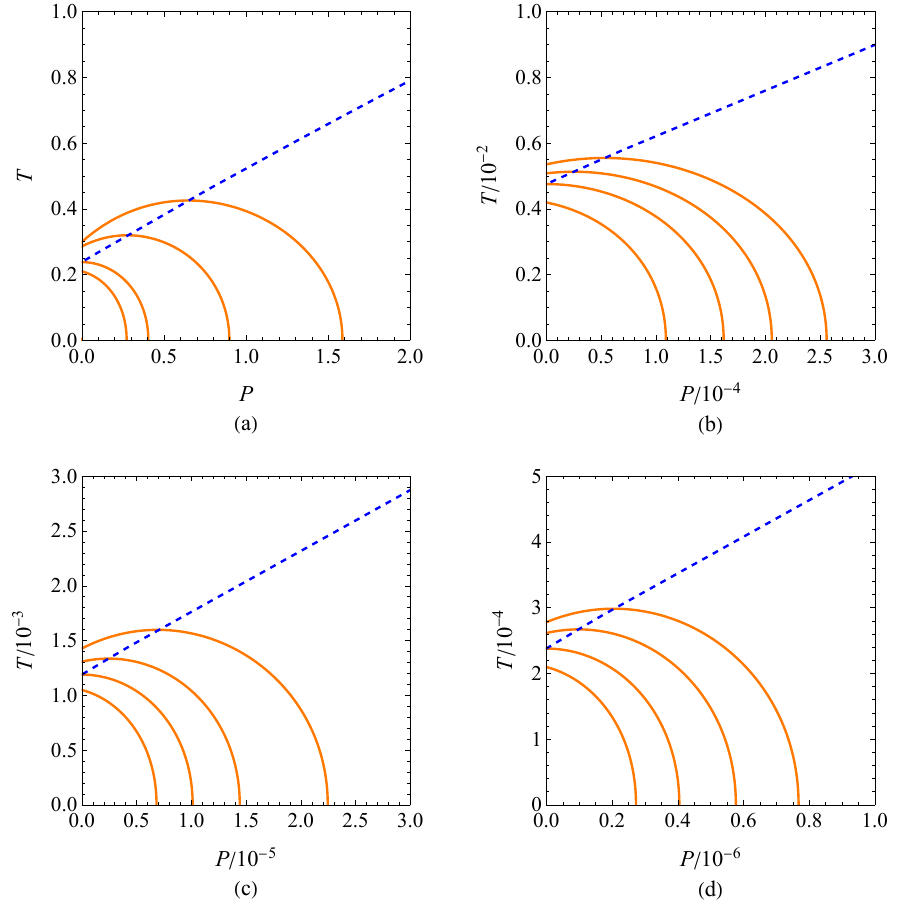}
	\caption{The constant mass curves (orange solid lines) and the inversion curves (blue dashed lines) for different parameters $a$. The constant mass curves expand outward as the mass increases. We set (a)$a=0.1,M=0.1,M_{i}^{min}(0.102),0.11,0.12$; (b)$a=5,M=5,M_{i}^{min}(5.1),5.2,5.3$; (c)$a=20,M=20,M_{i}^{min}(20.4),21,22$; (d)$a=100,M=100,M_{i}^{min}(102),105,108$.}\label{JT4}
\end{figure}

Fig.~\ref{JT4} shows the inversion curves and the families of Joule-Thomson expansion curves for different parameters $a$.

\section{Conclusion and outlook}\label{Sect5}
In this paper, we present a detailed and in-depth investigation of the thermodynamics of AdS BH with a Gaussian distribution within the framework of noncommutative geometry. Our study demonstrates that the Gaussian-distributed BH exhibits thermodynamic properties remarkably similar to those of AdS BH with a Lorentzian distribution in the same geometric setting. These similarities are specifically manifested in the modified first law of thermodynamics, the small BH-large BH phase transition, the $P-v$ critical phenomena, identical critical exponents, isobaric heat capacity, the zeroth-order phase transition, and the Joule-Thomson expansion.

A slight distinction from the Lorentzian source lies in the fact that Gaussian-distributed BH possesses a critical ratio (0.46531) that is evidently greater than 0.375 (the value corresponding to Van der Waals system), whereas the numerical value for the Lorentzian source (0.36671) is only slightly less than 0.375. Moreover, compared to the Lorentzian case, the zeroth-order phase transition effect in Gaussian-distributed BH is exceedingly subtle, with the relative increase in the Gibbs free energy being on the order of $10^{-3}\!\sim\!\!10^{-2}$. Consequently, this deviation is challenging to discern clearly from the $G\left(T\right)$ curve of the BH.

In the framework of noncommutative geometry, the Lorentzian and Gaussian distributions serve as two crucial alternative models for mass point distributions, both of which have been extensively discussed in the academic literature. Here, we present the key thermodynamic variables of the two types of BHs for comparison and reference (see Tab.~\ref{tab}).
\begin{table}[htbp]
\centering
\caption{Important thermodynamic variables of AdS BHs in noncommutative geometry. The result for BH with Lorentzian distribution is cited in Ref.~\cite{lorentzianBH}.}
\setlength{\tabcolsep}{4.26mm}{
	\begin{tabular}{ccc}
	\hline
	\hline
	Thermodynamic variables&Gaussian distribution&\quad Lorentzian distribution\\
	\hline
	Critical pressure $P_{c}$&$9.3973\times10^{-4}\Theta^{-1}$&\quad$1.3419\times10^{-4}\Theta^{-1}$\\
	Critical radius $r_{c}=v_{c}/2$&$6.0437\sqrt{\Theta}$&\quad$11.020\sqrt{\Theta}$\\
	Critical temperature $T_{c}$&$2.4412\times10^{-2}\Theta^{-\frac{1}{2}}$&\quad$8.0653\times10^{-3}\Theta^{-\frac{1}{2}}$\\
	Critical ratio $P_{c}v_{c}/T_{c}$&0.46531&\quad0.36671\\
	Minimum inversion temperature $T_{i}^{min}$&$1.1879\times10^{-2}\Theta^{-\frac{1}{2}}$&\quad$4.7016\times10^{-3}\Theta^{-\frac{1}{2}}$\\
	Minimum inversion mass $M_{i}^{min}$&$2.0441\sqrt{\Theta}$&\quad$4.7016\sqrt{\Theta}$\\ \hline
	\hline
	\end{tabular}}
	\label{tab}
\end{table}

Currently, there has been no work that systematically and comprehensively investigates the thermodynamics of AdS BH with a Gaussian distribution in the framework of noncommutative geometry. Our study aims to fill this gap. We sincerely hope that this article will enrich both the theory of noncommutative geometry and the research on BH thermodynamics in extended phase space, thereby providing a valuable reference for future studies.

Regarding future work, investigating charged BHs within noncommutative geometry and comparing the results with those of classical RN-AdS BH would be a highly valuable endeavor. This subject could further deepen our understanding of the influence of noncommutative geometry on BH thermodynamics. Furthermore, regarding the region of the noncommutative parameter, although $\sqrt{\Theta}$ is generally assumed to be on the order of the Planck length $\ell_{p}$, no universally accepted numerical value has been established. One approach is to constrain the noncommutative parameter using classical general relativity experiments within the solar system. However, the constraint obtained by this method depends on the choice of the distribution. Since the case of the Lorentzian distribution has already been computed~\cite{solartest}, we hope that future work will fill the gap by providing a parameter constraint using the Gaussian source.

\section*{Conflicts of interest}
The authors declare that there are no conflicts of interest regarding the publication of this paper.

\section*{Acknowledgments}
We want to thank School of Physical Science and Technology, Lanzhou University.

\section*{Data availability}
This work is a theoretical study. No data was used or generated during this research.

\bibliography{paper}

\end{document}